\documentclass[amstex,aps,preprint,nofootinbib,floatfix]{revtex4}%
\usepackage{graphicx}
\usepackage{amsmath}
\usepackage{bm}%
\usepackage{amsfonts}%
\usepackage{amssymb}

\begin{document}

\title{Are Bohmian trajectories real? \\
On the dynamical mismatch between
de Broglie-Bohm and classical dynamics in semiclassical systems}

\author{A.\ Matzkin}

\affiliation{Laboratoire de Spectrom\'{e}trie physique (CNRS
Unit\'{e} 5588), Universit\'{e} Joseph-Fourier Grenoble-1, BP 87,
38402 Saint-Martin, France}

\author{V. Nurock}

\affiliation{D\'{e}partement de Philosophie, UFR LLPhi,
Universit\'{e} de Paris-X-Nanterre, 200 Avenue de la R\'{e}publique,
92001 Nanterre, France}
\begin{abstract}
The de Broglie-Bohm interpretation of quantum mechanics aims to
give a realist description of quantum phenomena in terms of the
motion of point-like particles following well-defined
trajectories. This work is concerned by the de Broglie-Bohm
account of the properties of semiclassical systems. Semiclassical
systems are quantum systems that display the manifestation of
classical trajectories: the wavefunction and the observable
properties of such systems depend on the trajectories of the
classical counterpart of the quantum system. For example the
quantum properties have a regular or disordered aspect depending
on whether the underlying classical system has regular or chaotic
dynamics. In contrast, Bohmian trajectories in semiclassical
systems have little in common with the trajectories of the
classical counterpart, creating a dynamical mismatch relative to
the quantum-classical correspondence visible in these systems. Our
aim is to describe this mismatch (explicit illustrations are
given), explain its origin, and examine some of the consequences
on the status of Bohmian trajectories in semiclassical systems. We
argue in particular that semiclassical systems put stronger
constraints on the empirical acceptability and plausibility of
Bohmian trajectories because the usual arguments given to dismiss
the mismatch between the classical and the de Broglie-Bohm motions
are weakened by the occurrence of classical trajectories in the
quantum wavefunction of such systems.
\end{abstract}

\pacs {01.70.+w, 03.65.Ta, 03.65.Sq}

 \maketitle

\section*{\uppercase{1. Introduction}}

The de Broglie-Bohm (BB) causal theory of motion is an alternative
formulation of standard quantum mechanics (QM). Based on the seminal
ideas put forward by de Broglie (1927) and Bohm (1952), it is
probably the alternative interpretation that has been developed to
the largest extent, allowing to recover many predictions of QM while
delivering an interpretative framework in terms of point-like
particles guided by objectively existing waves along deterministic
individual trajectories. As put by Holland the aim is to develop a
theory of individual material systems which describes ''an objective
process engaged in by a material system possessing its own
properties through which the appearances (the results of successive
measurements) are continuously and causally connected'' (
Holland (1993) p.\ 17). Bohm and Hiley
(1985) state that embracing their interpretation shows ''there is no
need for a break or 'cut' in the way we regard reality between
quantum and classical levels''. Indeed one of the main advantages of
adopting the BB interpretative framework concerns the ontological
continuity between the quantum and the classical world: the
trajectories followed by the particles are to be regarded as real,
in the same sense that macroscopic objects move along classical
trajectories: ''there is no mismatch between Bohm's ontology and the
classical one regarding the existence of trajectories and the
objective existence
of actual particles'' (
Cushing 1994, p. 52). From a philosophical stand this ontological
continuity allows the de Broglie-Bohm interpretation to stand as a
realist construal of quantum phenomena, whereas from a physical
viewpoint the existence of trajectories leads to a possible
unification of the classical and quantum worlds (allowing for
example to define chaos in quantum mechanics). As is very
well-known, both points are deemed unattainable within standard
quantum mechanics: QM stands as the authoritative paradigm put
forward to promote anti-realism (not only in physics but also in
science and beyond
(Norris 1999)), whereas the emergence of the classical world from
quantum mechanics is still an unsolved intricate problem.

Concurrently, intensive investigations have been done in the last
twenty years on quantum systems displaying the fingerprints of
classical trajectories. Indeed in certain dynamical circumstances,
known as the \emph{semiclassical regime}, the properties of excited
quantum systems are seen to depend on certain properties of the
corresponding classical system. The recent surge of semiclassical
physics 
(Brack and Bhaduri 2003) has been aimed at studying nonseparable
systems in solid-state, nuclear or atomic physics that are hard to
solve or impossible to interpret within standard quantum mechanics
based on the Schr\"{o}dinger equation. One consequence of these
investigations was to increase the content of the quantum-classical
correspondence by highlighting new links between a quantum system
and its classical counterpart, such as the distribution of the
energy levels, related to\ the global phase-space properties of the
classical system. In particular classical chaos was seen to possess
specific signatures in quantum systems 
(Haake 2001). Moreover due to their high degree of excitation, some
semiclassical systems may extend over spatial regions of almost
macroscopic size.

In this work, we will be concerned by the de Broglie-Bohm account
of the properties of semiclassical systems. Contrarily to
elementary expectations, BB trajectories in semiclassical systems
have nothing in common with the trajectories of the corresponding
classical problem. This creates a mismatch between the BB account
of semiclassical systems and the one that is rooted in the
quantum-classical correspondence afforded by the semiclassical
interpretation. Our aim is to describe this mismatch, explain its
origin, and examine some of the consequences on the status of
Bohmian trajectories in semiclassical systems. Our starting point
will consist in a brief review of the salient features of BB
theory, insisting on the advantages of the interpretation relative
to standard QM (Sec. 2). We will then give a pedagogical
introduction to semiclassical physics and discuss the meaning of
the semiclassical interpretation (Sec. 3), insisting on how the
\emph{waves }of the quantum system depend on the
\emph{trajectories} of the corresponding classical system. In
particular if the classical motion is regular, the quantum
wavefunction and properties will be seen to reflect this
regularity, whereas chaotic classical motion translates quantum
mechanically into disordered wavefunctions and properties. These
features will be illustrated on a definite semiclassical system,
the hydrogen atom in a magnetic field. Sec.\ 4 will be devoted to
the exposition and discussion of BB\ trajectories for
semiclassical systems. We will explain why Bohmian trajectories
are necessarily highly nonclassical in this regime and examine the
consequences on the quantum-classical correspondence arising for
semiclassical systems in the context of quantum chaos. We will
complete our enquiry by assessing the specific problems that arise
from the dynamical mismatch between BB and classical trajectories
if the de Broglie-Bohm approach is intended to depict a real
construal of quantum phenomena in semiclassical systems. In
particular we will try to argue that semiclassical systems put
stronger constraints on the empirical acceptability and
plausibility of Bohmian trajectories because the usual arguments
given to justify the nonclassical behaviour of the trajectories
are weakened by the occurrence of classical properties in the
wavefunction of such systems.

Let us bring here two precisions.\ The first concerns the
terminology: throughout the paper we will indistinctively employ de
Broglie Bohm (BB) theory or Bohmian mechanics (BM) and related
expressions (such as Bohmian particle etc.) as strictly synonymous
terms referring to the theory summarized in Sec.\ 2, which presents
the mainstream version of the interpretation.\ We will therefore
disregard particular variations of the interpretation giving a
different ontological status to the the wavefunction, configuration
space, etc. The second precision concerns the validity of Bohmian
mechanics: let us state once and for all that we will only deal in
this work with the nonrelativistic theory, which as far as the
predictions are concerned is strictly equivalent to standard quantum
mechanics.\ Therefore our subsequent discussion will only deal with
the status and physical properties of the theoretical entities put
forward by BM, and does not touch upon the validity of the
predictions made by the theory.

\section*{\uppercase{2. Waves and particles in Bohmian mechanics}}

\subsection*{2.1 Basic formalism}

We briefly summarize the main features of the nonrelativistic de Broglie-Bohm
formalism, in its most commonly given form. The formalism starts from the same
theoretical terms that are encountered in standard QM, but makes the following
specific assumptions: the state $\psi$, solution of the Schr\"{o}dinger
equation, is given a privileged representation in configuration space. In
addition the $N$ particles that comprise the system are assumed to have at
every point of space (our usual space-time) a definite position and velocity.
The law of motion follows from the action of the ''pilot-wave'' $\psi.$ The
wave $\psi(x_{1}...x_{N}),$ where the $x_{i}$ are the positions of the
particles, is seen as a complex-valued field, a real physical field in a space
of dimension $3N$. The guiding law arises by employing the polar
decomposition
\begin{equation}
\psi(x,t)=\rho(x,t)\exp(i\sigma(x,t)/\hbar) \label{e1}%
\end{equation}
where $\rho$ and $\sigma$ are real functions that may depend on time. We now
restrict the discussion to a single particle of mass $m$ moving in a potential
$V(x,t)$ . The Schr\"{o}dinger equation becomes equivalent to the coupled
equations%
\begin{align}
\frac{\partial\sigma}{\partial t}+\frac{(\triangledown\sigma)^{2}}%
{2m}-\frac{\hbar^{2}}{2m}\frac{\triangledown^{2}\rho}{\rho}+V  &
=0\label{e3}\\
\frac{\partial\rho^{2}}{\partial t}+\triangledown(\frac{\rho^{2}%
\triangledown\sigma}{m})=0  &  . \label{e4}%
\end{align}
The first equation determines the Bohmian trajectory of the particle via the
relation%
\begin{equation}
p(x,t)=mv(x,t)\equiv\triangledown\sigma(x,t) \label{e5}%
\end{equation}
where $v$ is the velocity of the particle and $p(x,t)$ the associated momentum
field. It is apparent from Eq. (\ref{e3}) that the motion is not only
determined by the potential $V$ but also by the term%
\begin{equation}
Q(x,t)\equiv-\frac{\hbar^{2}}{2m}\frac{\triangledown^{2}\rho}{\rho},
\label{e6}%
\end{equation}
which for this reason is named the quantum potential. Indeed, in Newtonian
form, the law of motion takes the form%
\begin{equation}
m\frac{dv}{dt}=-\triangledown(V\left[  x(t),t\right]  +Q\left[  x(t),t\right]
). \label{e7}%
\end{equation}
In order to obtain a single trajectory Eq. (\ref{e5}) must be complemented
with the initial position $x(t=0)=x_{0}$ of the particle.

The main characteristics of the Bohmian trajectories follow from the
properties of the quantum potential. First $Q$ depends on the wave $\psi$ (and
more specifically on its form, not on its intensity). This implies that the
local motion of a given particle depends on the quantum state of the
\emph{entire} system (e.g.\ the properties -- mass, charge,... of all the
particles comprising the system, including their interactions), thereby
introducing nonlocal effects. Second the presence of $Q$ in Eq. (\ref{e7})
radically modifies the trajectory that would be obtained with the sole
potential $V$. For example a particle can be accelerated though no classical
force is present (as in free motion $V=0$). Conversely the quantum potential
may cancel $V$, yielding no acceleration where acceleration of the particle
would be expected on classical grounds. This is the case when the wavefunction
is real (e.g. for many stationary states), since the polar decomposition of
$\psi$ commands in this case that $\sigma$ vanishes. These points will be
illustrated and further discussed in Sec.\ 4.

Contact with standard quantum mechanics implies that $\rho$, which is the
amplitude of the physically real field $\psi$, gives the probability
amplitude, and hence $\rho^{2}$ gives the particle distribution in the sense
of statistical ensembles: Eq. (\ref{e4}) is a statement of the conservation of
the probability flow. Therefore the initial position $x_{0}$ lies somewhere
within the initial particle distribution $\rho_{0}^{2},$ but the precise
position of an individual particle is not known: indeed, the predictions made
by Bohmian mechanics do not go beyond those of standard quantum mechanics. But
the statistical predictions of QM are restated in terms of the deterministic
motion of a particle whose initial position is statistically distributed, this
ensemble distribution being in turn determined by $\psi$. The mean values of
quantum mechanical observables are in this way identified with the average
values of a statistical ensemble of particles.

\subsection*{2.2 Advantages of the interpretation}

Postulating the existence of specific trajectories followed by
point-like particles has no practical consequences as far as
physical predictions are concerned. The additional assumptions
introduced by the de Broglie-Bohm interpretation aim rather at
bridging the classical and the quantum worlds. This bridge,
underpinned by the coherent ontological package furnished by BM,
supports two interrelated levels: the first level concerns the
extension of scientific realism to the quantum world; the second
level addresses the unification of the classical and quantum
phenomena, thereby allowing to solve (in a conceptual sense) a
long-standing physical problem.

The difficulties of conceiving a scientific realist interpretation
of quantum phenomena are well-established
(see for example the celebrated paper by Putnam\footnote{In a recent
article Putnam (2005) reconsiders the problems raised by quantum
mechanics even for a broad and liberal version of scientific
realism, concluding on the possibility that "we will just fail to
find a scientific realist interpretation [of quantum mechanics]
which is acceptable". It is noteworthy that the de Broglie-Bohm
interpretation which was dismissed in Putnam (1965) on the ground
that the quantum potential has properties incompatible with realism
is considered as a possible realist interpretation by Putnam (2005)
on the basis of the hydrodynamic type of explanations allowed by BM.
As we will argue in Secs. 4 and 5, the hydrodynamic picture is at
the source of the difficulties encountered by BM in explaining the
properties of semiclassical systems.} (1965)), and standard QM in
the Copenhagen framework openly advocates instrumentalist and
operationalist approaches of
the theory. According to Bohm and Hiley (1985)
, the main motivation in introducing their interpretation is
precisely that ''it avoids making the distinction between realism in
the classical level and some kind of nonrealism in the quantum
level''. This is afforded by the ontological continuity that follows
from positing the existence of particles moving along deterministic
trajectories to account for quantum phenomena. In turn, this move
allows epistemological categories that are assumed to be necessary
for understanding 'what is really going on' (such as causality and
continuity) to operate in the realm of quantum mechanics.

The emergence of classical physics from quantum mechanics is still today one
of the main unsolved problems in physics. The ontological package furnished by
BM may give the key to the conceptual unification of quantum and classical
phenomena: the particles are the objects that are recorded in the experiments
and their existence is necessary ''so that the classical ontology of the
macroworld emerges smoothly without abrupt conceptual discontinuity''
(
Home 1997, p. 165). In a conceptual sense this allows to solve some
of the deepest quantum mysteries, such as the measurement problem
(Maudlin 1995) or the appearance of chaos in classical mechanics
from a quantum chaos substrate defined in terms of Bohmian
trajectories
(Cushing 2000).

Thus beyond the empirical equivalence between standard QM\ and the
de Broglie-Bohm approach, the latter's advantage is its declared
ability to offer a ''conceptually different view of physical
phenomena in which there is an objective reality whose existence
does not depend upon the observer'' 
(Cushing 1996 p. 6). The trajectories followed by a Bohmian particle
must then be taken as a realist construal of the properties of
quantum phenomena; Bohmian dynamics in semiclassical systems will be
investigated in this perspective.

\section*{\uppercase{3. Quantum systems displaying the fingerprints of classical
trajectories: the semiclassical regime}}

\subsection*{3.1 Reinforcing the quantum-classical correspondence}

The manifestation of classical orbits in quantum systems can take
many forms.\ Some examples (see
Brack and Bhaduri (2003) for reference to the original papers)
include: the recurrence spectra of excited atoms in external
fields that display peaks at times correlating with closed
classical trajectories; electron transport in nanostructures such
as quantum dots that show fluctuations correlating with the
periodic orbits that exist in a classical billiard having the same
geometry as the nanostructure; shell effects in nuclear fission
ruled by the fission paths computed in the phase-space of the
corresponding classical system. From a theoretical viewpoint the
origin of such manifestations lies on the validity of the
semiclassical approximation. As will be reviewed below the
semiclassical approximation is a framework that allows the
computation of quantum mechanical quantities from the properties
of the classical trajectories of the corresponding classical
system.

As a computational scheme, the semiclassical approximation is neutral
regarding the meaning of the computed quantum quantities. However the
semiclassical \emph{interpretation}, firmly grounded on the semiclassical
approximation goes further by explicitly linking the dynamical behaviour of a
quantum system to the behaviour and properties of the corresponding classical
system. The surge of semiclassical physics in the last 20 years
(see Gutzwiller (1990) or Brack and Bhaduri (2003)) has at least as
much to do with the explanatory success afforded by the
semiclassical interpretation than with improvements made in
numerical aspects of the semiclassical approximation (which have
been very important however).\ Indeed the semiclassical
interpretation relates universal properties of quantum systems to
the global phase-space typology of the underlying classical systems:
hence quantum systems whose classical counterpart is classically
chaotic universally possess certain signatures (such as the
statistical properties of the spectrum) very different from quantum
systems having a classically integrable counterpart. We give an
overview of the semiclassical approximation and then discuss the
dynamical explanation of the quantum-classical correspondence
allowed by the semiclassical interpretation.\ We will take as an
example a real system, the hydrogen atom in a magnetic field, for
which the Bohmian trajectories will be discussed in Sec.\ 4.

\subsection*{3.2 The semiclassical approximation}

The semiclassical approximation expresses the main dynamical
quantities of quantum mechanics in terms of classical theoretical
entities. This approximation is valid when the Planck constant
$\hbar$ is small relative to the classical action of the system. The
size of the action is roughly given by the product of the momentum
and the distance of a typical motion of the system (the action of
course grows as the system becomes bigger). The rest of this
paragraph gives explicitly the basic formulae of the semiclassical
approximation and its technical content is not essential to the
arguments developed in the rest of the paper.

The most transparent route to the derivation of the semiclassical
approximation starts from the path integral representation of the exact
quantum mechanical time-evolution operator, which for a single particle can
take the well-known form
(Grosche and Steiner 1998)
\begin{equation}
K(x_{2},x_{1};t_{2}-t_{1})=\int_{x_{1}}^{x_{2}}\mathcal{D}x(t)\exp
\frac{i}{\hbar}\left[  \int_{t_{1}}^{t_{2}}\left(  \frac{m}{2}\dot{x}%
^{2}-V(x)\right)  dt\right]  . \label{e12}%
\end{equation}
This expression propagates the probability amplitude from $x_{1}$ to
$x_{2}$ by considering all the paths that can be possibly taken
between these two points in configuration space, where $D$ is the
dimension. The term between square brackets is the classical action
$R(x_{2},x_{1};t_{2}-t_{1})$. When $R$ is much larger than $\hbar$
the integral in Eq. (\ref{e12}) can be approximately evaluated by
the method of stationary phase. The stationary points of $R$ are
simply the \emph{classical} paths connecting $x_{1}$ to $x_{2}$ in
the time $t_{2}-t_{1}$ and the propagator $K$ becomes approximated
(Chap. 5 of
Grosche and Steiner (1998)) by the semiclassical propagator $K^{sc}$%
\begin{equation}
K^{sc}(x_{2},x_{1};t_{2}-t_{1})=\left(  2i\pi\hbar\right)  ^{-D/2}\sum
_{k}\left|  \det\frac{-\partial^{2}R_{k}}{\partial x_{2}\partial x_{1}%
}\right|  ^{1/2}\exp\frac{i}{\hbar}\left[  R_{k}(x_{2},x_{1};t_{2}-t_{1}%
)-\phi_{k}\right]  . \label{e10}%
\end{equation}
The sum runs only on the \emph{classical} paths $k$ connecting
$x_{1}$ and $x_{2}$, and although all the quantities appearing in
this equation are classical except for $\hbar$, this expression
has a standard quantum mechanical meaning: in the semiclassical
approximation, the wave propagates only along the classical paths,
taking all of them simultaneously with a certain probability
amplitude. The weight of this probability amplitude depends on the
determinant in Eq. (\ref{e10}), which gives the classicaldensity
of paths (it is the inverse of the Jacobi field familiar in the
classical calculus of variations). $R_{k}$ is the classical action
along the trajectory $k;$ it satisfies the Hamilton-Jacobi
equation of classical mechanics
(Goldstein 1980)
\begin{equation}
\frac{\partial R(x,t)}{\partial t}+\frac{(\triangledown R(x,t))^{2}}%
{2m}+V(x)=0. \label{e13}%
\end{equation}
$\phi_{k}$ is an additional phase that keeps track of the points where the
Jacobi field vanishes.

Since usually most of the properties of quantum mechanical systems are
obtained through the eigenstates and eigenvalues of the Hamiltonian, it is
convenient to obtain a relevant semiclassical approximation in the energy
domain. The Green's function (i.e., the resolvent of the Hamiltonian) is
defined through the Fourier transform of the propagator $K$. The semiclassical
approximation is found by Fourier transforming $K^{sc},$ yielding%
\begin{equation}
G^{sc}(x_{2},x_{1};E)=2\pi\left(  2i\pi\hbar\right)  ^{-(D+1)/2}\sum
_{k}\left|  \frac{1}{\dot{x}_{1}^{\Vert}\dot{x}_{2}^{\Vert}}\det
\frac{\partial^{2}S_{k}}{\partial x_{2}^{\bot}\partial x_{1}^{\bot}}\right|
^{1/2}\exp i\left[  S_{k}(x_{2},x_{1};E)/\hbar-\phi_{k}\right]  . \label{e14}%
\end{equation}
$S_{k}(x_{2},x_{1};E)$ is the reduced action, also known as
Hamilton's characteristic function of classical mechanics
(Goldstein 1980), obtained by integrating the classical momentum
$p(x;E)$ along the classical trajectory $k$ linking $x_{1}$ to
$x_{2}$ at constant energy $E$. $\phi_{k}$ is a phase slightly
different than the one appearing in $K^{sc}$. The Green's function
is obtained by superposing all the classical paths $k$ at a fixed
energy $E$, each of them contributing according to the weight which
is related to the classical probability density of trajectories
(measuring how a pencil of nearby trajectories deviate from one
another). The classical probability density is obtained by squaring
the term  $\left| {\hspace{.8cm} }\right|^{1/2}  $, here written in
terms of coordinates parallel $(\Vert)$ and perpendicular $(\bot)$
to the motion along the periodic orbit.

The most useful quantity that is obtained from the resolvent $G$ is the level
density $d(E)=\sum_{n}\delta(E-E_{n})$ which quantum mechanically gives the
energy spectrum by peaking at the eigenvalues $E_{n}$. $d(E)$ is obtained by
taking the trace of $G$. Taking the trace of $G^{sc}$ yields the semiclassical
approximation to the level density%
\begin{equation}
d^{sc}(E)=\bar{d}(E)+\sum_{j\in\mathrm{po}}A_{j}(E)\cos\left(  \frac{S_{j}%
(E)}{\hbar}-\phi_{j}\right)  , \label{e16}%
\end{equation}
known as Gutzwiller's trace formula
(Gutzwiller 1990). The sum over $j$ now runs only on the
\emph{closed classical periodic orbits} that exist at energy $E$
irrespective of their starting point. $S_{j}$ is the reduced action
accumulated along the $j$th periodic orbit,%
\begin{equation}
S_{j}(E)=\oint_{\mathcal{C}_{j}}p(x;E)dx
\end{equation}
and $A_{j}(E)$ is the amplitude depending on the classical period $T_{j}$ of
the orbit and on its monodromy matrix (giving the divergence properties of the
neighboring trajectories, such as the Lyapunov exponents). Finally $\bar
{d}(E)$ is the mean level density, proportional to the volume of the classical
energy shell in phase-space i.e. the points enclosed in the surface
$H(p,x)=E$, $H$ being the classical Hamiltonian. $\bar{d}$ varies smoothly
with $E$ and cannot contribute to the peaks in the level density, which are
solely due to the contribution of the periodic orbits. Other spectral
quantities, such as the cumulative level density that are employed when
studying the distribution of the energy levels are also obtained in terms of
the classical periodic orbits from $d^{sc}(E)$.

Eqs. (\ref{e13}), (\ref{e14}) and (\ref{e16}) unambiguously relate
the evolution and spectral properties of the quantum system to the
classical trajectories of the corresponding classical system. These
approximations are expected to be valid for quantum systems in the
semiclassical regime.

\subsection*{3.3 Semiclassical systems in quantum mechanics}

\subsubsection*{3.3.1 Quantum dynamics depending on classical phase-space properties}

Risking a tautology, we will say a quantum mechanical system to be
semiclassical if the semiclassical approximation holds. The main
requirement is thus that the exact quantum-mechanical propagator can
take the approximate form given by Eq. (\ref{e10}). This happens
when the classical action $R$ is large relative to $\hbar$. The
number of systems amenable to a semiclassical treatment is huge and
we refer to the field textbooks
(Gutzwiller 1990, Brack and Bhaduri 2003) where many examples can be
found.\ Semiclassical systems are generally very excited (implying
high energies and large average motions, of almost macroscopic
sizes), rendering quantum computations delicate to undertake.\ The
semiclassical approximation is sometimes the only available
quantitative tool.\ Even when exact quantum computations are
feasible, the solutions of the Schr\"{o}dinger equation do not give
any clue whatsoever on the dynamics of the system. The r\^{o}le of
the semiclassical interpretation is then to explain the dynamics of
the system, relying on the association between the properties of the
quantum system and the structure of classical phase-space. This
increased content of the quantum-classical correspondence, sometimes
known as ''quantum chaos'', involves both average and individual
properties of the corresponding classical system.

In classical mechanics, it is well known
(Arnold 1989) that trajectories in \emph{integrable systems} are
confined in phase-space to a torus; projected in configuration space
these trajectories have a regular motion, behaving in an orderly
manner.\ On the other hand typical trajectories in a \emph{chaotic
system} explore all of available phase-space and have an
unpredictable behaviour (in the sense that two nearby trajectories
diverge exponentially in time).\ In semiclassical systems, the
individual energy eigenfunctions are either organized around the
torii when the corresponding classical system is integrable, or
scattered throughout configuration space when the classical system
is chaotic. In the former case the existence of classical torii
translate quantum mechanically into the dependence of the energy on
integer (sometimes called 'good') quantum numbers, which count the
number of periodic windings around the torus of the periodic orbits
(this is why the quantum numbers increase by one at each winding).
When the corresponding classical system is chaotic there are no more
'good' quantum numbers. There still are periodic orbits, and their
classical properties (amplitude and action) determine (from Eq.
(\ref{e16})) the position of the quantized energies.

Individual classical periodic orbits play a r\^{o}le in explaining
quantum phenomena such as recurrences or scars. Recurrences in
time are the product of the periodic partial reformation of the
time evolving wavefunction in configuration space ('revival') and
are related to the large scale fluctuations of the energy
spectrum. The recurrence times happen at the period of the
periodic orbits of the corresponding classical system [Eq.
(\ref{e10})]. Scars concern an increase of the probability density
of the wavefunctions along the periodic orbits of the
corresponding classical system. Other properties of a
semiclassical quantum system depend on the average properties of
classical trajectories, such as the typical region of phase-space
explored by a trajectory.\ These average properties are reflected
in the statistical distribution of the energy eigenstates of the
quantum system (average sum rules involving the mean behaviour of
the periodic orbits yield different statistical distributions
according to whether the corresponding classical dynamics is
integrable or chaotic; see e.g. Berry 1991) .

\subsubsection*{3.3.2 Illustrative example: the hydrogen atom in a magnetic field}

\begin{figure}[tbh]
\begin{center}
\includegraphics[height=6.6in,width=4.in]{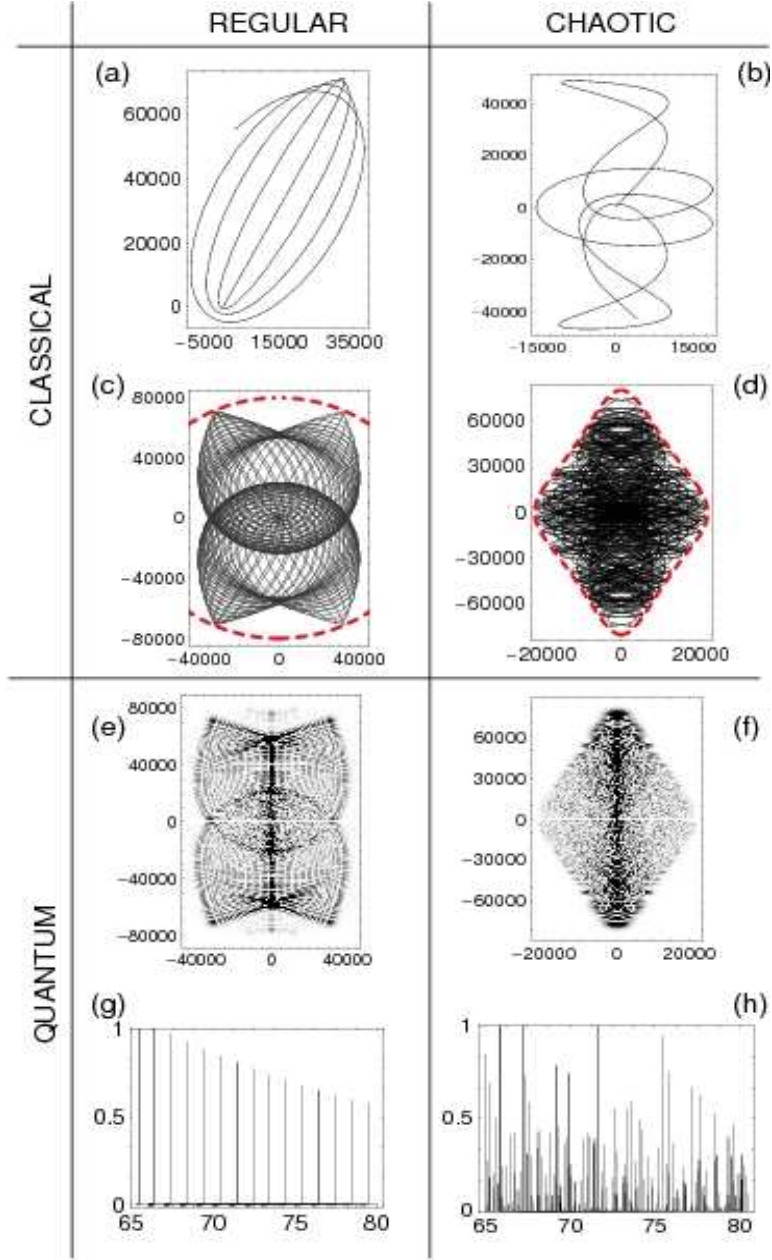}
\caption[]{Illustration of the quantum-classical correspondence for
the hydrogen atom in a magnetic field. The left column shows
features when the dynamics of the classical system is
\emph{regular}, the right column when the dynamics is
\emph{chaotic}. (a)-(d) shows \emph{classical trajectories} for the
electron. (e)-(f) displays \emph{wavefunctions} obtained from
quantum computations for identical dynamical regimes [(e) as in (c),
(f) as in (d)]. (g)-(h) gives an experimentally observable quantity
(here obtained from quantum computations) by keeping again the
dynamical regimes identical as above. See text (Sec. 3.3.2) for more
details. } \label{fig1}
\end{center}
\end{figure}

We will illustrate how the semiclassical interpretation works in practice by
resorting to a specific example, the hydrogen atom in a magnetic field. This
system has been thoroughly investigated both theoretically and experimentally
and the semiclassical approximation is known to hold. Moreover from a de
Broglie-Bohm standpoint, Bohmian trajectories for this system have been
recently obtained
(Matzkin 2006). Since an illustration involving the relation
between spectral eigenvalue statistics and average classical
properties would necessarily be very technical, we shall limit our
examples to some global qualitative aspects and the r\^{o}le of
the shortest classical periodic orbits, illustrated in Figs. 1 to
3.

The hydrogen atom in a uniform magnetic field $B$ is an effective
nonseparable problem in two dimensions (due to cylindrical symmetry;
for details the interested reader is referred to the review paper
by Friedrich and Wintgen (1989)). The electron is subjected to the
competing attractive Coulomb and magnetic fields. The dynamics of
the classical system does not depend separately on the electron's
energy $E$ and the field strength $B,$ but on the scaled energy
defined by the ratio $\epsilon=EB^{-2/3}$ (this property is due to a
scaling invariance). When $\epsilon\rightarrow-\infty$ (in practice
for $\epsilon \lesssim-0.7$) the Coulomb field dominates and the
dynamics is regular (near-integrable regime). As the relative
strength of the magnetic field increases, the classical dynamics
turns progressively chaotic: a mixed phase-space situation holds for
$-0.7\lesssim\epsilon\lesssim-0.2$ and for $\epsilon\gtrsim-0.15$
the phase space is fully chaotic (Poincar\'{e} surfaces of section
are given in Friedrich and Wintgen (1989)). The scaling property
also holds for the quantum problem, and one can therefore compute
wavefunctions corresponding to a fixed value of $\epsilon.$

Fig.\ 1\ encapsulates the nature of the quantum-classical
correspondence for semiclassical systems. (a)-(d) show trajectories
of the electron for the \emph{classical} hydrogen in a magnetic
field problem. $\varrho$ is the horizontal axis, $z$ the vertical
axis (the magnetic field is in the $+z$ direction). The nucleus is
fixed at $\varrho=0$, $z=0$. The scale is given in atomic units
($5.3\times 10^{-11}$ m), so that the trajectory spans a rather
large distance for a microsystem (the electron goes as far as  0.01
mm from the nucleus). In (a)-(b), the electron is launched from the
nucleus and evolves for a short time along the trajectory. The
initial conditions are the same in (a) and (b), but the dynamical
regime differs: (a) shows the trajectory when the dynamics of the
classical system is regular ($\epsilon=-1$), and (b) when the
dynamics is chaotic ($\epsilon=-0.15$). In (c) [resp. (d)] the
trajectory shown in (a) [resp. (b)] evolves for a long time; in each
case we have also plotted the symmetric trajectory (e.g. (c) shows
(a) and the trajectory symmetric to (a) starting downward). The
dashed line indicates the bounds of the region accessible to the
electron at the given energy. When the dynamics is regular the
trajectory evolves in a regular manner and occupies only a part of
the accessible region in configuration space (this is due to the
fact that in \emph{phase-space} the trajectory is confined to a
torus). When the dynamics is chaotic [(d)] the trajectory evolves
disorderly and occupies ergodically all of the available region.
(e)-(h) display features for the \emph{quantum} hydrogen in a
magnetic field problem. In (e) we have shown the density-plot of a
wavefunction for the same dynamical regime (identical value of
$\epsilon$) as the trajectory shown in (c) (the wavefunction was
obtained by solving numerically the Schr\"{o}dinger equation). The
resemblance with (c) is explained semiclassically from the
quantization of the torus explored by the classical trajectory. The
wavefunction has a regular aspect (for example in the organization
of the nodes). (d) gives the plot of the wavefunction when the
corresponding classical system is in the chaotic regime as in (d).
The wavefunction has a disordered aspect when compared to (e), and
has an overall shape quite similar to the shape formed by letting a
classical trajectory evolve as in (d). Finally, (g) and (h) show an
experimentally observable quantity, the photoabsorption spectrum,
that gives a rough idea of how the energy levels are distributed
(the horizontal axis gives the energy of the level in terms of a
number $\nu$ with $E=-1/2 \nu ^2$). Regular dynamics gives a
photoabsorption spectrum with a regular aspect, characterized by
evenly spaced lines (again a consequence of torus quantization which
guarantees the existence of 'good' quantum numbers). When the
corresponding classical dynamics is chaotic, the spectrum [shown in
(h)] has a complex aspect, reflecting the disappearance of an
ordered structure in the underlying classical phase-space.

\begin{figure}[tb]
\begin{center}
\includegraphics[height=2.1in,width=3in]{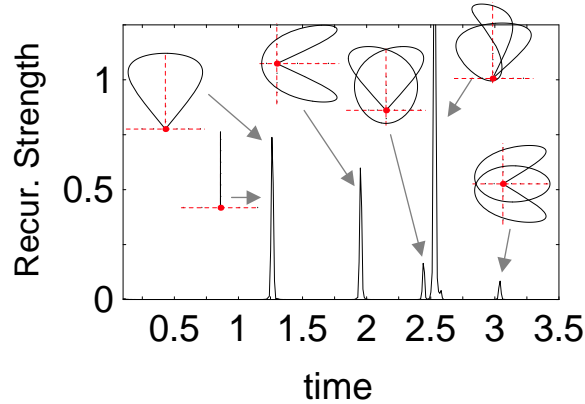}
\caption[]{Recurrence spectrum of the hydrogen atom in a magnetic
field obtained from quantum computations (the time is given in
scaled units). The diagrams show the shape of the orbits closed at
the nucleus of the corresponding (scaled) classical problem in the
$(\varrho,z)$ plane ($B$ is along the vertical $z$ axis). The arrows
indicate the orbit whose period matches the time of a given peak in
the recurrence spectrum. Adapted from Matzkin (2006). } \label{fig2}
\end{center}
\end{figure}

Fig.\ 2 shows a recurrence spectrum, giving the part of the initial
wavefunction that returns to the nucleus as a function of time, the
electron, initially near the nucleus, being excited at $t=0$. This
spectrum results from a quantum computation but the same type of
spectra has been observed experimentally\footnote{Note that by using
Eqs. (\ref{e10})-(\ref{e14}), the semiclassical approximation allows
to compute quantitatively a recurrence spectrum almost identical to
the exact one shown on the figure, obtained by solving the
Schr\"{o}dinger equation. This confirms we are dealing with a system
in the semiclassical regime.}
(Main et al 1994). The peaks appear at times matching the period of
the classical periodic orbits shown on the figure next to the
peaks.\ The interpretative picture is the following: once the
electron gets excited (e.g. by a laser), the wavefunction propagates
in configuration space along the classical trajectories. Therefore
the peaks in the recurrence spectrum indicate that the part of the
wavefunction that returns to the nucleus does so by following the
classical periodic orbits closed at the nucleus. Several orbits that
have the same or nearly the same period can contribute to a given
peak.\ In that case the phases $\phi_{k}$ of Eq. (\ref{e10}) which
rule the interference between those orbits are crucial so that the
semiclassical computations reproduce the exact height of the peak.

As a final illustration, Fig. 3 shows the localization of an energy
eigenfunction on a periodic orbit of the classical problem. The left
panel shows the wavefunction of an energy eigenstate when the
corresponding classical dynamics is mixed (part of phase space is
chaotic, part regular); the wavefunction was obtained by numerically
solving the Schr\"{o}dinger equation. The probability density is
seen to occupy all of the available diamond-shaped region (as in
Fig. 1(d)), but the striking feature is the very strong density
concentrated on an spring-like shape. This shape is a periodic orbit
of the classical system existing at the same energy. The periodic
orbit, obtained by solving the classical equations of motion, is
drawn on the right panel.
\begin{figure}[tb]
\begin{center}
\includegraphics[height=2.1in,width=3.7in]{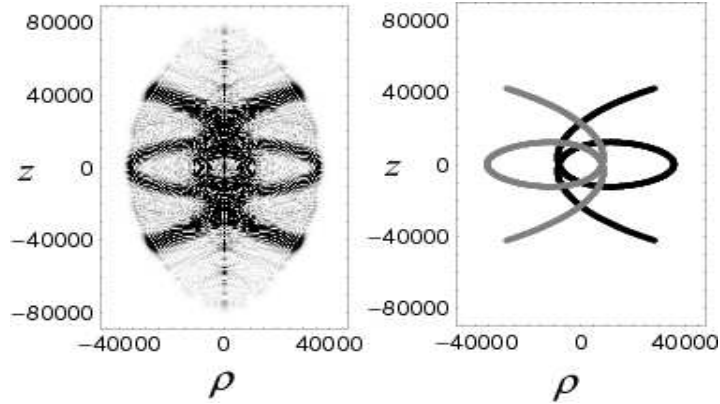}
\caption[]{Left: Wavefunction of an excited energy eigenstate of the
hydrogen atom in a magnetic field. Right: Periodic orbit of the
corresponding classical system at the same energy (actually the
periodic orbit [in black] is plotted along with its partner,
symmetric by reflection on the $z$ axis [in gray]). } \label{fig3}
\end{center}
\end{figure}

\subsection*{3.4 Status of the semiclassical interpretation}

As we have introduced it, the semiclassical interpretation is an
asymptotic approximation to the Feynman path integral. Whereas the
exact path integral involves all the paths connecting two\ points in
spacetime, the semiclassical approximation only takes into account
the classical paths. It seems unwarranted to require more from the
approximation than what is found in the exact expression, namely the
propagation of a wave in configuration space by taking
simultaneously all the available paths. No point particle can be
attached to the path integral trajectories -- it is actually an
instance of a simultaneous sum over paths regarding the propagation
of the wave. By itself the semiclassical interpretation cannnot and
does not aim at explaining the emergence of classical mechanics from
a quantum substrate. What does emerge are the structural properties
(shape, stability) of the classical trajectories. In this sense, the
small $\hbar$ condition (in relative terms) that defines the
semiclassical regime appears as a necessary (but definitely not
sufficient) ingredient in accounting for the classical domain. The
semiclassical interpretation tells us that there are classical
orbits in the wavefunction, but does not aim at unraveling what the
wavefunction becomes in the classical world.\ Note that the same can
be said regarding classical mechanics in the Hamilton-Jacobi
formalism: this formalism contains the same classical theoretical
entities that are employed in the semiclassical approximation.\
Obviously classical mechanics does not contain any traces of
periodic wave propagation, but the wavefront of the classical action
does propagate like a shock wave in configuration space (see Sec.\
10-8 of
Goldstein (1980)). As a theoretical entity the action is by itself a
non referring epistemic term and it is only by making additional
assumptions that the motion of an ensemble of trajectories can be
extracted from the propagating wavefront of the action, recovering
the ontology of classical mechanics in Newton's form.

We point out notwithstanding that the semiclassical interpretation
has also been considered as a theory in its own right, distinct from
classical but also from quantum mechanics
(Batterman 1993, 2002). Such an assessment has been made on the
basis of the different nature of the explanations afforded by the
semiclassical interpretation relative to the standard quantum
theory: Batterman argues that emergent properties are characteristic
of asymptotic theories, as these cannot be reduced to the
fundamental theories that lack the conceptual resources necessary to
the interpretation of asymptotic phenomena. Without necessarily
endorsing this viewpoint, the semiclassical interpretation
nevertheless attributes to the classical concepts it employs similar
virtues. The fact that one ignores what the quantum-mechanical
theoretical entities refer to explains why the putatively
fundamental theory (quantum mechanics) is unable to account for the
emergence of classical mechanics, and why in semiclassical physics
the interpretative framework relies essentially on classical
conceptual resources. Hence in semiclassical physics one speaks of
the quantum properties as being ''determined'' by the properties of
the underlying classical system, or on the wavefunction
''depending'' on the classical trajectories. It must be remembered
however that stricto-sensu, the semiclassical interpretation only
establishes an enlarged, precise and universal correspondence
between the properties of quantum systems in the semiclassical
approximation and those of their classical equivalents.

\section*{\uppercase{4. Bohmian trajectories in the semiclassical
regime}}

\subsection*{4.1 From quantum to classical trajectories}

As mentioned in Sec.\ 2, one of the main advantages of the de
Broglie-Bohm interpretation is that BM allows to bridge the quantum
and classical worlds, by way of a quantum point-like particle
following a precise trajectory\footnote{Note that a complete account
of the emergence of the classical world should also take care of the
fate of the pilot-wave as the classical limit is approached. We will
leave this aspect of the problem out of the scope of the present
work, essentially because as we have just noticed, the semiclassical
interpretation is not concerned by this problem. Moreover the
ontological status of the sole wavefunction (i.e., without a
particle) in Bohmian mechanics is not very different from what is
proposed by other interpretations that assume the objective
existence of the wavefunction (Zeh 1999).}. Of course, trajectories
in the quantum domain are generically nonclassical, due to the
presence of the quantum potential. This quantum state dependent
potential enters the equations for the Bohmian trajectories in Eq.
(\ref{e3}); without this term Eq. (\ref{e3}) would become the
classical Hamilton-Jacobi equation (\ref{e13}). The presence of the
quantum potential term in Eq. (\ref{e3}) leads to highly
nonclassical solutions even for intuitively simple systems.\ The
best-known example is probably the particle in a box problem, which
raised Einstein's well-know criticism of the de Broglie-Bohm
interpretation (a thorough discussion can be found in Sec.\ 6.5\ of
Holland 1993): classically a particle in a box would follow a to and
fro motion, its constant velocity being reversed when the particle
hits the boundary of the box. According to BM however there is no
particle motion, because the quantum potential cancels the classical
kinetic energy. The same behaviour arises for stationary
wavefunctions like the energy eigenstates: for example the electron
in a hydrogen atom is either at rest (if the azimuthal quantum
number is 0) or it displays an asymmetric motion around the
quantization axis. On the other hand the classical trajectories for
that system (the familiar ellipses of planetary motion) need to
fulfil the symmetry of the classical potential (just as the
wavefunction does), which is broken by the quantum potential term.

Now having Bohmian trajectories in the quantum domain different
from the trajectories in the classical domain is not necessarily a
problem. The problem is that as the classical world is approached,
there are no physical criteria that will unambiguously make the
quantum potential vanish and lead to classical trajectories,
irrespective of how the classical limit is defined. Different
opinions to circumvent this important difficulty from a physical
standpoint have been given (these arguments will be developed and
discussed in Sec. 5). One possibility is that some quantum
systems, or specific states thereof (such as the energy
eigenstates) simply do not have a classical limit
(Holland 1993); conversely some classical systems may not be
obtained as limiting cases of an underlying quantum problem,
entailing that BM and classical mechanics would have exclusive
(albeit partially overlapping) domains of validity
(Holland 1996, Cushing 2000). Another argument asserts that since
closed quantum systems cannot be observed in principle (since they
always interact with a measurement apparatus, an environment etc.),
'apparent' trajectories (resulting from the measurement interaction)
are bound to be different from the 'real' ones (Bohm and Hiley 1993,
Ch. 8). But if trajectories could be inferred without perturbing a
quantum system, the ones predicted by BM would be found, not the
classical ones
(Bohm and Hiley 1985). More recently tentative proposals of
recovering classical trajectories from Bohmian mechanics in simple
systems by combining specific types of wavepackets and environmental
decoherence arising from interactions with the environnement have
been put forward (Appleby 1999b, Bowman 2005), although these
treatments lack sufficient generality.
In this context, the specificity of semiclassical systems follows
from the fact that they are closed quantum systems that
nevertheless display the manifestations of classical trajectories.
It is thus instructive to examine the behaviour of Bohmian
trajectories in semiclassical systems.

\subsection*{4.2 Bohmian trajectories and semiclassical wave-propagation}

It is straightforward to make the case that in semiclassical systems
Bohmian trajectories are highly nonclassical, as in any quantum
system. Indeed in the de Broglie-Bohm interpretation, the velocity
field given by Eq. (\ref{e5}) is proportional to the quantum
mechanical probability
density current $j(x,t)=\hbar\psi^{\ast}(x,t)\overleftrightarrow{\nabla}%
\psi(x,t)/2mi$ through $j=\rho^{2}v$ (recall $v(x,t)$ is the
velocity of the Bohmian particle).\ Since frome Eq. (\ref{e10}) an
evolving
wavefunction takes the form%
\begin{equation}
\psi(x,t)=\sum_{k}A_{k}(x,x_{0},t)\exp i(R_{k}(x,x_{0},t)/\hbar-\phi_{k}),
\label{18}%
\end{equation}
where $A_{k}$ includes the determinant of the propagator and
quantities relative to the initial wavefunction, the probability
current $j(x,t)$ at some point $x$ will be given by a double sum
containing interference terms: the net probability density current
arises from the interference of several classical trajectories taken
simultaneously by the wave, as required by the path integral
formulation. Therefore except in the exceptional case in which there
is a single classical trajectory (that the probability current must
therefore follow), the net probability density guiding the Bohmian
particle will not flow along one of the classical trajectories that
act as backbones of the wavefunction in the semiclassical regime.
The particle in a box case gives a particularly simple
illustration.\ At a fixed energy, there are only 2 classical
trajectories passing through a given point, one in each direction,
so that the net probability flow is 0, translating in BM as no
motion for the particle.

\begin{figure}[tb]
\begin{center}
\includegraphics[height=2.1in,width=5.4in]{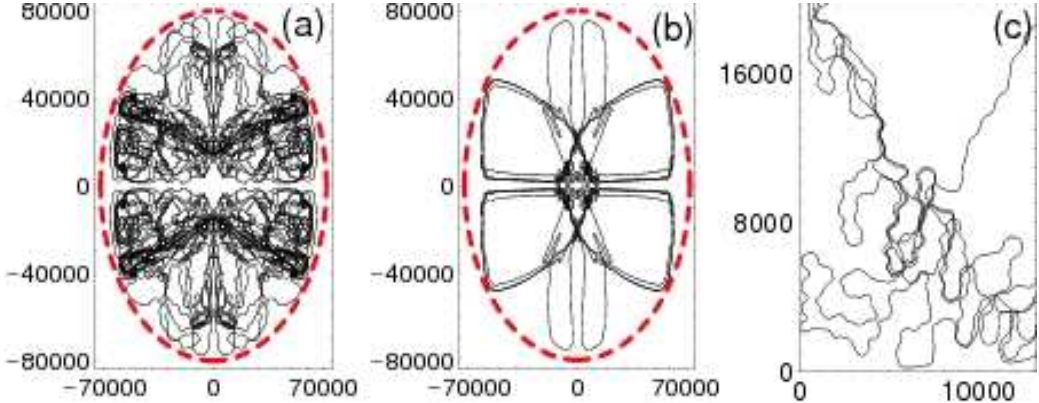}
\caption[]{Bohmian trajectories for the hydrogen in a magnetic
field problem in the \emph{regular} regime of Fig. 1 (classical
regular trajectories, ordered quantum properties). (a) and (b)
each contain a single trajectory in the $\varrho>0$, $z>0$
quadrant and 3 symmetric replicates of this trajectory in the
other quadrants (see text). The initial position of the Bohmian
electron is the same in (a) and (b), but the initial quantum state
is different. (c) shows a zoom of (b) in the region near the
nucleus (for the positive quadrant only). The dashed lines
indicate the bounds of the classically accessible region, as in
Fig. 1.} \label{fig4}
\end{center}
\end{figure}

For the hydrogen atom in a magnetic field problem examined in Sec.\
3, de Broglie-Bohm trajectories can be computed and compared to the
classical ones (Matzkin 2006).\ Typical Bohmian trajectories for the
electron are shown in Figs.\ 4 and 5, for the dynamics in the
regimes shown in Fig.\ 1. Fig. 4 shows trajectories corresponding to
the regular column in Fig. 1 ($\epsilon=-1$ corresponding to
classical regular trajectories and quantum properties having a
regular aspect). The sole difference between Fig. 4(a) and (b)
concerns the choice of the initial wavefunction (which contains more
eigenstates in case (a)). It is important to stress that a Bohmian
trajectory cannot cross either the $\varrho$ or $z$
axis\footnote{The entire $\varrho$ axis is a node on which the
quantum potential becomes infinite (and hence cannot be crossed).
When approaching the $z$ axis the velocity of a Bohmian particle
goes to zero, as there is no net density current through this axis,
and is then reversed (hence the axis is not crossed).}. Figs. 4(a)
and (b) contain each a Bohmian trajectory in the $\varrho>0$, $z>0$
quadrant and three symmetric copies of this trajectory in the
remaining quadrants (in agreement with the symmetry of the
statistical distribution). The trajectory in Fig. 4(a) has a chaotic
aspect and occupies most of configuration space, whereas the
trajectory in Fig. 4(b) has a more regular aspect, as the Bohmian
particle retraces several times a similar figure. However when this
trajectory is zoomed in the region near the nucleus (Fig 4(c); we
only show the 'original' trajectory in the positive quadrant), the
regular aspect is not evident. We therefore see that the
correspondence between the classical dynamical regime and quantum
properties encapsulated in Fig. 1 is not obeyed by BB trajectories.
The converse is also true: Fig. 5 shows a Bohmian trajectory (and
its 3 symmetric replicates) in the chaotic dynamical regime of the
right column of Fig. 1 (classically chaotic trajectories and
disordered quantum properties). The Bohmian trajectory is visibly
regular, in the sense that the particle retraces the same regions of
space in a similar fashion as can be seen in the zoom, Fig. 5(b),
and occupies only a small part of the region of configuration space
available to the particle.

\begin{figure}[tb]
\begin{center}
\includegraphics[height=2.1in,width=3.7in]{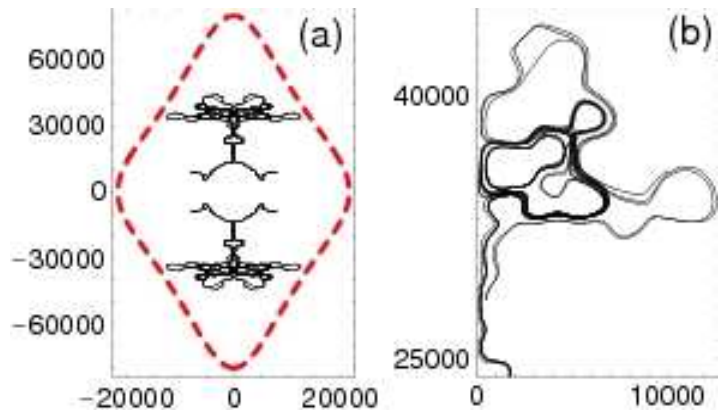}
\caption[]{Same as Fig. 4 but in the \emph{chaotic} regime of Fig. 1
(classical chaotic trajectories, disordered quantum properties). (a)
shows a Bohmian trajectory in the positive quadrant and its 3
replicates, (b) zooms in a specific region.} \label{fig5}
\end{center}
\end{figure}

Another difference in the behaviour of de Broglie-Bohm trajectories
with regard to the quantum-classical correspondence for
semiclassical systems can be seen in the energy eigenstates. The
eigenstates are organized and sometimes localized along the periodic
orbits of \ the classical hydrogen in a magnetic field problem as
seen in Fig.\ 3, but a Bohmian particle in an energy eigenstate has
no motion in the $(\varrho,z)$ plane (it has no motion at all if
$m=0,$ or orbits around the $z$ axis so that the trajectory remains
still in the $(\varrho,z)$ plane if $m\neq0$). We also note as
another consequence of the nonclassical nature of the BB
trajectories that the periodic recurrences of the type shown in
Fig.\ 2, which appear at times matching the periods of return of
classical closed orbits (and their repetitions), cannot be produced
by a an individual Bohmian trajectory. Indeed Bohmian trajectories
do not possess the classical periodicities visible in the peaks, and
an ensemble of different Bohmian trajectories compatible with a
given statistical distribution is necessary to account for the
recurrences
(Matzkin 2006). This is a straightforward consequence stemming
from the fact that a Bohmian particle moves along the streamlines
of the probability flow. Then the evolution of the wavefunction
between the initial and the recurrence times can only be obtained
if the complete ensemble of streamlines is taken into account
(Holland 2005).

\subsection*{4.3 Quantum chaos and the quantum-classical correspondence}

We have given examples of de Broglie-Bohm trajectories for the
hydrogen atom in a magnetic field problem and seen that their
features are unrelated to the properties of the underlying classical
system and therefore do not fit in the quantum-classical
correspondence scheme arising from the semiclassical interpretation.
This is of course a general statement: analogue results were
obtained for square and circular billiards, which are classically
integrable systems, but where Bohmian trajectories were found to be
either regular or chaotic, depending on the choice of the initial
wavefunction (the initial state that also gives the initial
statistical distribution)
(Alcantara-Bonfim et al 1998). In right triangular billiards chaotic
or regular Bohmian trajectories were found for the \emph{same}
initial distribution but different initial position of the particle
(de Sales and Florencio 2003). In one of the earliest studies of the
Bohmian approach to quantum chaos
(Parmenter and Valentine 1995), a two dimensional uncoupled
anisotropic harmonic oscillator (a separable system having a
deceptively simple regular classical dynamics) was shown to display
chaotic Bohmian trajectories (see also
Efthymiopoulos and Contopoulos 2006).

The chaotic nature of Bohmian trajectories is due to the
state-dependent quantum potential, given by Eq. (\ref{e6}).\ In
particular when $\rho(x,t)$ vanishes, the quantum potential
becomes singular.\ This happens at the nodes of the wavefunction.
Employing the hydrodynamic analogy (the probability density flow
carries the BB trajectories), the nodes correspond to vortices of
the probability fluid.\ It has recently been confirmed
(Wisniacki and Pujals 2005) that the vortices are at the origin of
the generic chaotic behavior of Bohmian trajectories; these authors
obtained chaotic trajectories even in an isotropic harmonic
oscillator, the 'most regular' classical integrable system.
Conversely since as we have just mentioned a Bohmian particle has no
motion in an eigenstate, it is always possible to obtain a regular
Bohmian trajectory (even for a disordered quantum system
corresponding to classically chaotic dynamics) by taking an initial
distribution composed of a 2 or 3 eigenstates  with a very strong
weight for one of these states.

Thus, from the perspective developed in this paper, it is clear
that Bohmian mechanics spoils the quantum-classical correspondence
that arises in the semiclassical regime, in the sense that de
Broglie-Bohm trajectories can be chaotic or regular irrespective
of the dynamical characteristics of the corresponding classical
system which are in correspondence with observable quantum
properties. As we have emphasized above, the energy eigenstates in
quantum semiclassical systems are organized around the phase-space
structures of the corresponding classical system (Fig. 2).\ The
distribution of the energy levels is also directly dependent on
the underlying classical dynamics in two ways: a universal
relation valid for any system, depending on the mean properties of
the periodic orbits, and a system specific behaviour, depending on
individual periodic orbits\footnote{Then the shortest periodic
orbits play a prominent r\^{o}le, e.g. in systems where the
semiclassical wave diffracts on obstacles having a specific
geometry, as in
Matzkin and Monteiro (2004).}. In the de Broglie Bohm approach,
the trajectories are entirely determined by the precise form of
the quantum-mechanical wavefunction: there is no manner in which
the topology of the trajectories can account for the structural
aspects of the wavefunction. This is why quantum chaos in Bohmian
terms and quantum chaos understood in the semiclassical sense are
radically divergent. The former strives to define chaos exactly as
in classical mechanics, by examining the properties of quantum
trajectories; but Bohmian trajectories will necessarily remain
unrelated to the properties of the corresponding classical system,
and will be unable to explain the manifestations of chaotic
classical trajectories in quantum systems. On the contrary,
quantum chaos in the semiclassical sense accounts for purely
quantum mechanical features by linking them to the dynamical
properties of the corresponding classical system, in particular to
its chaoticity; but as we have mentioned above, the semiclassical
interpretation has no more ontological ambition than what is found
in the path integral formulation of quantum mechanics.

\section*{\uppercase{5. Constraints on the empirical acceptability and reality of Bohmian trajectories in
semiclassical systems}}

From within a purely internal quantum approach governed by the
Schr\"{o}dinger equation, semiclassical systems have a peculiar
property: the dynamics of these quantum systems depends on the
trajectories of the corresponding classical system. This 'peculiar
property' arises naturally in the path integral formulation. On the
other hand, we have seen above that the de Broglie-Bohm
interpretation appears to contradict the enlarged version of the
quantum-classical correspondence stemming from the semiclassical
interpretation. From the standpoint of the benefits that should
emerge from adopting Bohmian mechanics (see Sec.\ 2), this may
appear as unexpected. Indeed, if BM is unable to account for the
presence of classical trajectories in semiclassical quantum systems,
how will the interpretation explain the emergence of macroscopic
classical trajectories? Taking the de Broglie-Bohm interpretation as
a realist construal of quantum phenomena, what is implied when
asserting with BM that the trajectories of the particles are highly
nonclassical in reality although the shape of the classical
trajectories is directly visible in the wavefunction? These
questions touch upon issues that have been examined in more general
contexts by supporters and critics of the interpretation, but these
issues take in semiclassical systems a particularly acute form. We
review the type of answers that have been given and examine their
implications regarding a de Broglie-Bohm account of semiclassical
systems.

A first argument involves a reassessment of the relation between
quantum and classical mechanics considered as fundamental physical
theories. We have seen above that a strong motivation for embracing
the de Broglie Bohm interpretation is that  BM  allows to fill the
gap between the quantum and the classical domains. BM would thus
provide ''an attractive understanding of the classical limit''
(Callender and Weingard 1997). Given that as far as the dynamics is
concerned Bohmian trajectories in closed systems are not classical,
this motivation is reassessed by putting the stress on the
ontological continuity while questioning the necessity that
classical mechanics needs to emerge from the quantum substrate.\
This is why Holland
(1996) suggests that quantum and classical mechanics should be
regarded as two different theories having a partial overlap: the
latter should not be expected to emerge from the former.\ This
move allows Holland to conclude that the fact that ''classical
dynamics is not generally a special case [of the de Broglie Bohm
approach] has then no implications for the validity of the
interpretation'' (p. 109). Hence the dynamical mismatch between BB
and classical trajectories cannot constitute an objection, each
theory having its own domain of validity. Applying this argument
to semiclassical systems is however not straightforward: these
systems are characterized by the manifestations of classical
trajectories, implying for the least a minimal partial overlap
that is not reflected in the dynamics of the Bohmian particle. To
stand by Holland's conclusion, one would need to assume that
semiclassical systems constitute an instance of non-overlap for
the dynamics of the particle (since the Bohmian trajectories are
nonclassical) despite having the quantum wavefunction organized
according to the underlying classical phase-space, and thus in
correspondence with the motion of the particle in the
corresponding classical system, thereby displaying a type of
overlap on a different level.

Arguments of the same kind have been made on quantum chaos, where
the non-overlap is now made between the aspect of the wavefunction
and the dynamics of the Bohmian particle. As concisely put by
Cushing
(2000) ''BM has certain conceptual and technical resources (not
available in standard QM) that allow it to address, and arguably to
resolve, two long-standing and difficult issues in quantum theory
(i.e., the classical limit and quantum chaos)''. In the de
Broglie-Bohm theory, these specific conceptual resources -- and
therefore the issues of quantum chaos and the classical limit --
involve the particle dynamics, not the wave. Commenting their
findings on the chaotic Bohmian trajectories obtained in a
classically integrable system, Parmenter and Valentine (1995) can
thus state that ''there is no contradiction between the fact that
the wave-function associated with a quantum mechanical state is not
chaotic, while at the same time a causal trajectory associated with
the state is chaotic''.
A contradiction can indeed be avoided but at the expense of an
account containing at least three levels of explanation: the
Bohmian dynamics level, determined by the wavefunction, which in
turn is determined, for semiclassical systems, by the underlying
classical phase-space. Hence regular classical dynamics is
reflected in a regular quantum wavefunction but not in the chaotic
Bohmian trajectories that are the ones followed by the particle
'in reality'. Therefore the necessary non-overlap between the wave
aspect and the dynamics of the Bohmian particle yields in
semiclassical systems another instance of the dynamical mismatch
between BB and classical trajectories, only the latter being
reflected in the aspect of the wavefunction. The same type of
explanation must be considered to explain the localization of the
energy eigenstates on the classical periodic orbits as in Fig. 3:
the wavefunction is organized around the classical periodic
orbits, the probability density of the wave reflecting precisely
the classical density distribution (i.e., the classical amplitude
of a pencil of classical trajectories). In the BB approach, this
means that the statistical distribution of Bohmian particles
depends on the properties of the classical periodic orbits present
in the wavefunction; but the motion of the Bohmian particles will
be unrelated to the classical motion. Again, it is consistent in
BB terms to dismiss any requirement of overlap between the
dynamics of the particle and the aspect of the wavefunction, but
in semiclassical systems this move implies a non-overlap with the
classical trajectories of the particle that actually determine the
shape of the wavefunction.

Although the non-overlap type of explanations may have an overdone
flavour when applied to semiclassical systems, it seems impossible
to avoid them if the de Broglie-Bohm interpretation is taken as a
realist account of the quantum mechanics of semiclassical systems.
The reason is the incompatibility between the hydrodynamic picture
underlying the motion of the BB particle along a \emph{single path}
connecting two spacetime points on the one hand, and the wave
picture where the wave takes simultaneously all the \emph{multiple
interfering paths} between the two points on the other. In the
hydrodynamic picture the motion of the particle depends entirely on
the local properties of the probability density flow. In the wave
picture the paths actually construct the wavefunction. Hence if
semiclassical wavefunctions are built from classical paths, a
Bohmian trajectory will only be sensitive to the necessarily
nonclassical flow. It is noteworthy that proponents of the de
Broglie Bohm interpretation tend to dismiss the path integral
approach as a mathematical tool without any substantial physical
implication (e.g. Sec.\ 6.9 of
Holland (1993)). The semiclassical systems discussed in this work
contradict this assertion since the classical dynamics is directly
reflected in the wavefunction, and observable quantities such as
recurrence spectra display in an unequivocal way the interfering
paths (see the example given in Fig. 2).\

A different argument of epistemological nature has often been put
forward to explain the irrelevance of the non-classical behaviour of
Bohmian trajectories to the quantum-classical transition problem.
Since BB trajectories in closed system are unobservable in principle
(they are hidden), no observable consequences can arise from their
non-classicality. When observed, the quantum system will be
interacting with an environment (such as a measurement apparatus).
The \emph{apparent} Bohmian trajectory of the open interacting
system is the one that should be compared to the classical
trajectory, not the\emph{ real} trajectory of the isolated system.
Hence the motion along an individual BB trajectory of a closed
system has no bearing on the empirical acceptability of the theory,
provided the statistical predictions of quantum mechanics are
recovered
(Appleby 1999a). In a general context, Fine
(1996) remarked that this type of argument employs the same
positivist lines of defense that are generally found in
vindication of the Copenhagen interpretation, namely that what
matters is the predictive agreement with experiments and that it
is impossible in principle to observe isolated, noninteracting
systems or trajectories. But departing from standard quantum
mechanics, which states that describing an isolated system has no
meaning, BM proposes a full description, in terms of waves and
particles trajectories, of the reality of an individual isolated
system. The specific difficulty arising for semiclassical systems
is that classical trajectories are present in the wavefunction of
the closed system. Since Bohmian mechanics regards the
wavefunction as a real field, the retreat behind the shield of the
sole empirical acceptability may appear as insufficient: in view
of the epistemological advantages associated with the de
Broglie-Bohm approach recalled in Sec. 2.2 (Cushing 1994), one
would expect a dynamical explanation for the presence of the
classical trajectories in the wavefunction (the statistical
distribution) while individual trajectories followed by the
Bohmian particle are all nonclassical. Furthermore, even if one
accepts that classical trajectories arise dynamically from
environmental interactions, the fact is that the noninteracting
semiclassical systems display several observable manifestations of
the classical trajectories.

The question then becomes under which conditions it is plausible
to assume that the \emph{real} (nonclassical) Bohmian trajectory
of the closed semiclassical system can be markedly different from
the (classical) \emph{apparent} trajectory of the measured
interacting system even though the pilot-wave of the closed system
is itself built on these classical trajectories, whose
manifestations are experimentally observed. The dynamical mismatch
induces a tension between the supposedly real dynamics of the
particles (following de Broglie-Bohm trajectories), the
statistical distribution of the particles (determined by the
underlying classical properties) and the classical trajectories
(observed by means of a physical interaction). In this respect, it
may be noted that the justification of the "special assumption"
(Holland 1993, p. 99) by which the guiding field $\psi$ gives at
the same time the statistical distribution of the particles
$\rho^{2}$ has always been a delicate point in the de Broglie-Bohm
approach\footnote{See
Durr et al (1996), where this postulate is dubbed the 'quantum
equilibrium hypothesis'. Bohm and Hiley (1993, Ch. 9) suggest that
this postulate can be explained in terms of the average
equilibrium resulting from an underlying stochastic motion.
Semiclassical systems would then call for more elaborate models to
account for the regular or disordered aspect of the wavefunction,
depending on the dynamics of the classical counterpart.}. We are\
not questioning the possibility that specific auxiliary
assumptions can be added to the 'special assumption', so that in
semiclassical systems the statistical distribution $\rho^{2}$ is
made to depend on the classical dynamics, but not the motion of
the Bohmian particle, except when interacting with a specific
environment, inducing the oberved classical trajectories. We
should however consider such auxiliaries relative to the
motivational advantages associated with the interpretation, viz.
the extension of realism to the domain of quantum phenomena and
the unification of classical and quantum mechanics. On both of
these points, semiclassical systems bring specific constraints
that must be met by Bohmian mechanics to ensure a credible account
of the dynamical properties of these systems. Regarding the first
point we will only mention here the realist demand for epistemic
constraints on the auxiliary assumptions so as to avoid ad-hoc
explanations and the ensuing underdetermination dilemmas, given
that the real dynamical behaviour of a quantum system cannot be at
the same time chaotic and regular\footnote{As a specific example,
consider for instance a semiclassical system presenting disordered
quantum properties, whose classical counterpart has chaotic
dynamics but whose BB dynamics is regular. The system cannot be
said to be chaotic or regular depending on how we decide to
\emph{choose} the ontology, on the ground that arbitrary
auxiliaries will always be able to entail the same observational
evidence (Devitt 2002). The interplay between the auxiliary
constraints and the types of underdetermination dilemmas arising
from the empirical equivalence of interpretations with
antagonistic ontological commitments in semiclassical systems will
be examined elsewhere in a forthcoming work.}.
As for the second point our account of the physics of semiclassical
systems suggests that classical mechanics appears in the quantum
world by structuring the wavefunctions, rather than by orchestrating
the streamlines of the probability flow. This statement must
actually hold to explain the correspondence between quantum and
classical systems illustrated in Fig. 1. Since semiclassical systems
constitute a bridge between the quantum and the classical worlds,
the status of Bohmian trajectories in such systems hinges on a\
better understanding of the relations between the hydrodynamic
picture and the path integral formulation of quantum phenomena, and
of a possibly elusive unification of both formulations.

\section*{\uppercase{6. Concluding remarks}}

Pauli was among the first to criticize both de Broglie's first
proposal of the pilot-wave (at the 1927\ Solvay conference) as well
as Bohm's rediscovery of it some 25 years later. In his contribution
to de Broglie's 60th birthday volume, Pauli (1952) dismissed the de
Broglie-Bohm approach as ''artificial metaphysics''
 because the BB approach breaks the
correspondence between classical mechanics and standard quantum
mechanics regarding the symmetric treatment of canonically conjugate
variables. Indeed, in quantum mechanics a Fourier transform links
the representation of the wavefunction in terms of one variable
(e.g. the position) to the representation in terms of its canonical
conjugate (the momentum). In the BB approach the position
representation is the only meaningful one; the momentum of the
Bohmian particle is given by the guidance equation, not by a Fourier
transform. Pauli's criticism is certainly not accidental: although
less well-known Pauli was the first to obtain the correct form of
the semiclassical propagator from the Feynman path integral
(Choquard and Steiner 1996). In the semiclassical approximation the
Fourier transform gives the purely classical relation between the
two action functions, one expressed in terms of one variable (the
position), the other in terms of its canonical conjugate (the
classical momentum of the particle).

It would appear that a formal objection favouring a correspondence
relation between theoretical entities in two different physical
theories (quantum and classical mechanics) challenges the
ontological continuity postulated by Bohmian mechanics, which is
grounded on a particle guided by the current density of the wave.
However, one of the conclusions of this work is that this objection
is not purely formal, because semiclassical systems do exist in
nature. We have seen that the consequences of this quantum-classical
correspondence are visible in semiclassical systems through the
manifestations of classical trajectories on the quantum observables.
The dynamics predicted in the de Broglie-Bohm approach, assumed to
represent the real motion of a quantum particle, conflicts with this
correspondence. Interestingly, the ontological continuity posited in
the de Broglie-Bohm interpretation as well as the quantum-classical
correspondence that transpires in semiclassical systems both go one
way, from the classical down to the quantum world (in terms of
classical-like particles pursuing nonclassical trajectories in the
first case, in terms of nonclassical waves moving on classical
trajectories in the latter). Yet in both cases, the missing piece of
the puzzle can only be put into place by finding the proper upward
way.

\vspace{1cm}

\end{document}